\documentclass[superscriptaddress, aps, prl, reprint, twocolumn, amsmath, amssymb, showpacs]{revtex4-1}
\usepackage{graphicx}  
\usepackage{mathptmx}
\usepackage{epstopdf}
\usepackage{dcolumn}   
\usepackage{bm}        
\usepackage{amssymb}   
\usepackage{amsmath}   
\usepackage{amsthm}
\usepackage{chemarrow}
\usepackage{color}
\usepackage{xcolor} 
\usepackage{mathrsfs}
\usepackage{float}
\usepackage{physics}
\usepackage{bbold}
\usepackage{bbm}
\usepackage[toc,page]{appendix}
\usepackage{longtable}
\usepackage{multirow}
\usepackage{upgreek}
\usepackage{epstopdf}
\usepackage{ulem}
\usepackage{graphicx}
\usepackage{lipsum}

\usepackage[]{ezedits}
\defineEdit{WJ}{\color[rgb]{0.5,0.3,0.9}}{\color[rgb]{0.5,0.3,0.9}}
\defineEdit{DB}{\color[rgb]{0,0.3,1}}{\color[rgb]{0,0.3,1}}

\usepackage[a4paper, colorlinks=true, linkcolor=blue, citecolor=blue,
pdfauthor={ },
pdftitle={ },
pdfsubject={ },
pdfkeywords={ }]{hyperref}
\usepackage[version=4]{mhchem}
\usepackage{ulem,fancyvrb}

\theoremstyle{plain}
\newtheorem*{theorem*}{Theorem}

\begin{document}
	

\title{New Constraints on Axion Mediated Dipole-Dipole Interactions}
 
\date{\today}
	
	\author{Zitong Xu}
	\affiliation{
		School of Instrumentation Science and Opto-electronics Engineering, Beihang University, Beijing, 100191, China}

    \author{Xing Heng}
	\affiliation{
		School of Instrumentation Science and Opto-electronics Engineering, Beihang University, Beijing, 100191, China}

	\author{Guoqing Tian}
	\affiliation{
		School of Instrumentation Science and Opto-electronics Engineering, Beihang University, Beijing, 100191, China}
  
    \author{Di Gong}
 	\affiliation{
		Hangzhou Extremely Weak Magnetic Field Major Science and Technology Infrastructure Research Institute,
Hangzhou, 310051, China}

	\author{Lei Cong}
	\affiliation{Johannes Gutenberg University, Mainz 55128, Germany}
	\affiliation{Helmholtz Institute Mainz, Mainz, 55128, Germany}

	\author{Wei Ji}
	\email[Corresponding author: ]{wei.ji.physics@gmail.com}
	\affiliation{Johannes Gutenberg University, Mainz 55128, Germany}
	\affiliation{Helmholtz Institute Mainz, Mainz, 55128, Germany}
    \affiliation{School of Physics and State Key Laboratory of Nuclear Physics and Technology, Peking University, Beijing 100871, China}

	\author{Dmitry Budker}
	\affiliation{Johannes Gutenberg University, Mainz 55128, Germany}
	\affiliation{Helmholtz-Institute, GSI Helmholtzzentrum fur Schwerionenforschung, Mainz 55128, Germany}
	\affiliation{Department of Physics, University of California at Berkeley, Berkeley, California 94720-7300, USA}

    \author{Kai Wei}
    \email[Corresponding author: ]{weikai@buaa.edu.cn}
    \affiliation{
	School of Instrumentation Science and Opto-electronics Engineering, Beihang University, Beijing, 100191, China}
 	\affiliation{
		Hangzhou Extremely Weak Magnetic Field Major Science and Technology Infrastructure Research Institute,
Hangzhou, 310051, China}

\begin{abstract} 
The search for axions sits at the intersection of solving critical problems in fundamental physics, including the strong CP problem in QCD, uncovering the nature of dark matter, and understanding the origin of the universe’s matter-antimatter asymmetry. 
The measurement of axion-mediated spin-dependent interactions offers a powerful approach for axion detection. However, it has long been restricted to regions outside the “axion window” due to a significant trade-off: the need to effectively suppress the magnetic leakage from highly polarized spin sources while simultaneously detecting sub-femtotesla level exotic physics signals at sub-decimeter-scale distances. In this work, we report new experimental results on axion-mediated exotic spin-spin interactions using an iron-shielded SmCo$_5$ spin source in combination with a specially designed self-compensation comagnetometer. Employing a composite shielding structure, we achieved a suppression of the magnetic field by up to $10^{11}$. This enabled us to establish new constraints on the coupling between electrons and neutrons, surpassing previous experimental limits by more than 10000 times within the axion window. Furthermore, we also set strongest constraints on the coupling between electrons and protons. 
The proposed method holds substantial potential not only for advancing the search for new physics beyond the Standard Model but also for enabling transformative applications in biological and chemical research.
\end{abstract}

\maketitle

The Standard Model of particle physics has been remarkably successful in explaining a wide range of phenomena but falls short in addressing several key issues, such as the nature of dark matter, dark energy, and the matter-antimatter asymmetry in the universe.  These unresolved issues strongly suggest the existence of new physics beyond the Standard Model. To bridge these gaps, various beyond-the-Standard Model theories have proposed the existence of new light bosons~\cite{Peccei:1977ur,graham2018spin}, with particular focus on spin-0 axion and axion-like particles (ALPs) (in this article, we refer to both axions and ALPs as ''axions" )~\cite{athron2021global,co2021predictions,bloch2020axion}.
Axions are candidate constituents of cold dark matter. The challenge in the search for axions is that their mass is unknown and can, in principle, be of any value with a range spanning many orders of magnitude.
Several theories constrain the most probable axion mass range to 0.01\,meV-1\,meV~\cite{turner1990windows,youdin1996limits,borsanyi2016calculation,klaer2017dark,ballesteros2017unifying}, which is the so called "axion window". However, previous laboratory haloscope experiments for axion dark matter research such as the
Axion Dark Matter experiment (ADMX)~\cite{PhysRevLett.120.151301} and astrophysical observations such as the SN1987A~\cite{Engel1990emission} and Sudbury Neutrino Observatory (SNO)~\cite{PhysRevLett.126.091601} primarily explored mass ranges outside the window~\cite{PhysRevLett.122.191302,bloch2020axion}. Recently, experimental efforts have shifted toward exploring the axion window, including haloscope experiments (Orpheus~\cite{PhysRevD.91.011701}, MADMAX~\cite{brun2019new}, ORGAN~\cite{Quiskamp_2022}, and CAPP~\cite{PhysRevLett.133.051802}). 

If axion exists, it could mediate exotic spin-dependent forces~\cite{moody1984new,dobrescu2006spin}, 
which has garnered considerable attention in recent years~\cite{safronova2018search,cong2024spin}. Searches for spin-dependent forces have the advantage of covering a wide range of the axion mass without scanning. A comprehensive overview of theoretical and experimental developments is available in a recent review \cite{cong2024spin}. Numerous experimental approaches are used or proposed to investigate these forces, including torsional pendulums and oscillators~\cite{aldaihan2017calculations}, atomic magnetometers~\cite{ji2023constraints,wei2022constraints,almasi2020new,kim2018experimental,wang2022limits} and nitrogen-vacancy  centers in diamond~\cite{jiao2021experimental}, nuclear magnetic resonance (NMR) ~\cite{su2021search,xu2024constraining} and other advanced technologies~\cite{yan2013new,stadnik2018improved,ren2021search}. However, there are fewer experiments detecting it within the axion window~\cite{wang2022limits,su2024new}.  The significant challenge is that measurements within the axion window demand that the distance between a sensitive detector and a high-spin-density source should be sub-decimeter.
Ideally, one could choose a high spin density source such as a SmCo$_5$ magnet and a high sensitivity atomic sensor to enhance the sensitivity to exotic interaction\,\cite{cong2024spin}.
However, simultaneously achieving suppression of usual magnetic field leakage, which is about the earth magnetic field, and ultrasensitive measurement of non-magnetic exotic signal ,which is sub-femtotesla level in this range is extremely challenging.

In this work, we tackle this challenge by employing a combination of advanced magnetic shielding and a self-compensation (SC) sensor. The SC sensor is designed to be insensitive to usual magnetic fields while retaining ultrahigh sensitivity to non-magnetic fields. We report a method that achieves a magnetic-field suppression factor of more than \(10^{11}\) with the help of a composite magnetic shield. On top of this, there is the suppression that comes from the use of a self-compensated magnetometer, measured to be 25. Compared with the  recent work to search for axion in the axion window \cite{wang2022limits}, the net spin in our spin source is more than 10\(^6\)-fold higher. 
Meanwhile, the sensitivity of our atomic sensor is also two orders of magnitude higher \cite{wei2023ultrasensitive}.
We apply this approach to search for exotic electron-neutron (or electron-proton) coupling in the  (0.12-1\,meV) range within the axion window. Having found no significant evidence for axion-mediated new force, we improve previous experimental constraints on electron-neutron coupling by over 10000 times within the axion window and set the strongest constraints on electron-proton coupling. The approach demonstrated in this study exhibits significant potential for a wide range of biological and chemical applications that rely on the combination of a high-field prepolarization region and a low-field detection region.

\textit{Exotic dipole-dipole interactions --} Among the various exotic interactions studied, specific terms take precedence based on the couplings under consideration, as detailed in Ref.\,\cite{cong2024spin}. In particular, the dipole-dipole term, which is the focus of this study, is significant because it provide the strongest constraints on the pseudoscalar-pseudoscalar couplings associated with axion\cite{cong2024spin}. The dipole-dipole interactions we measure in this experiment is

\begin{eqnarray}
V_{pp} &= -g_p^Xg_p^Y \frac{\hbar^3}{16\pi c}\frac{1}{m_Xm_Y}\left[\boldsymbol{\sigma}_X\cdot\boldsymbol\sigma_Y^{\,\prime} \left(\frac{1}{r^3}+\frac{1}{\lambda r^2}+\frac{4\pi}{3}\delta(\boldsymbol{r})\right)\right. \nonumber\\
&\left.-(\boldsymbol{\sigma}_X\cdot \hat{\boldsymbol{r}})(\boldsymbol\sigma_Y^{\,\prime}\cdot \hat{\boldsymbol{r}}) \left( \frac{3}{r^3} + \frac{3}{\lambda r^2} + \frac{1}{\lambda^2 r} \right)\right]e^{-{r}/{\lambda}} \, , 
\label{eq.v3}
\end{eqnarray}

\noindent where \(\hbar\) is the reduced Planck constant, \(c\) is the speed of light, \(\boldsymbol{\sigma}_{X}\) and \(\boldsymbol{\sigma}_Y^{\,\prime}\) are vectors of Pauli matrices representing the spins \(\boldsymbol{s}_i=\hbar \boldsymbol{\sigma}_i/2\) of the two fermions \(X\) and \(Y\), in our case, the electron and neutron/proton, \(m_{X}\) and \(m_{Y}\) are the corresponding masses, \(\lambda=\hbar/m_a c\) is the force range and $m_a$ is the axion mass, and \(r\) is the distance between two fermions. Here, we adopt the exotic potentials for macroscopic experiments as recommended in the review \cite{cong2024spin}, and the $\delta(\mathbf{r})$ term is zero in our case, but is non-negligible on atomic scale \cite{fadeev2022pseudovector}.  

\textit{Experimental setup --} The K-Rb-\(^{21}\)Ne comagnetometer in this experiment is similar to the one in Ref.\,\cite{wei2022constraints}. There is a 12\,mm-diameter spherical cell containing 3.5 amagats of \(^{21}\)Ne, 50\,torr N\(_2\) and K and Rb atoms with density ratio of about 1/100 at the center of the comagnetometer. The cell is heated to 190\,$^{\circ}$C. The K atoms are polarized with circularly polarized light tuned to the K D1 line along the \textit{z} axis. The Rb and \(^{21}\)Ne atoms are spin polarized by spin exchange collisions. Hybrid optical pumping is utilized to improve the polarization uniformity. The precession of Rb atoms is detected with linearly polarized probe light propagating along \textit{x}-axis.

\begin{figure}[htb]
    \centering
\includegraphics[width=0.99\linewidth]{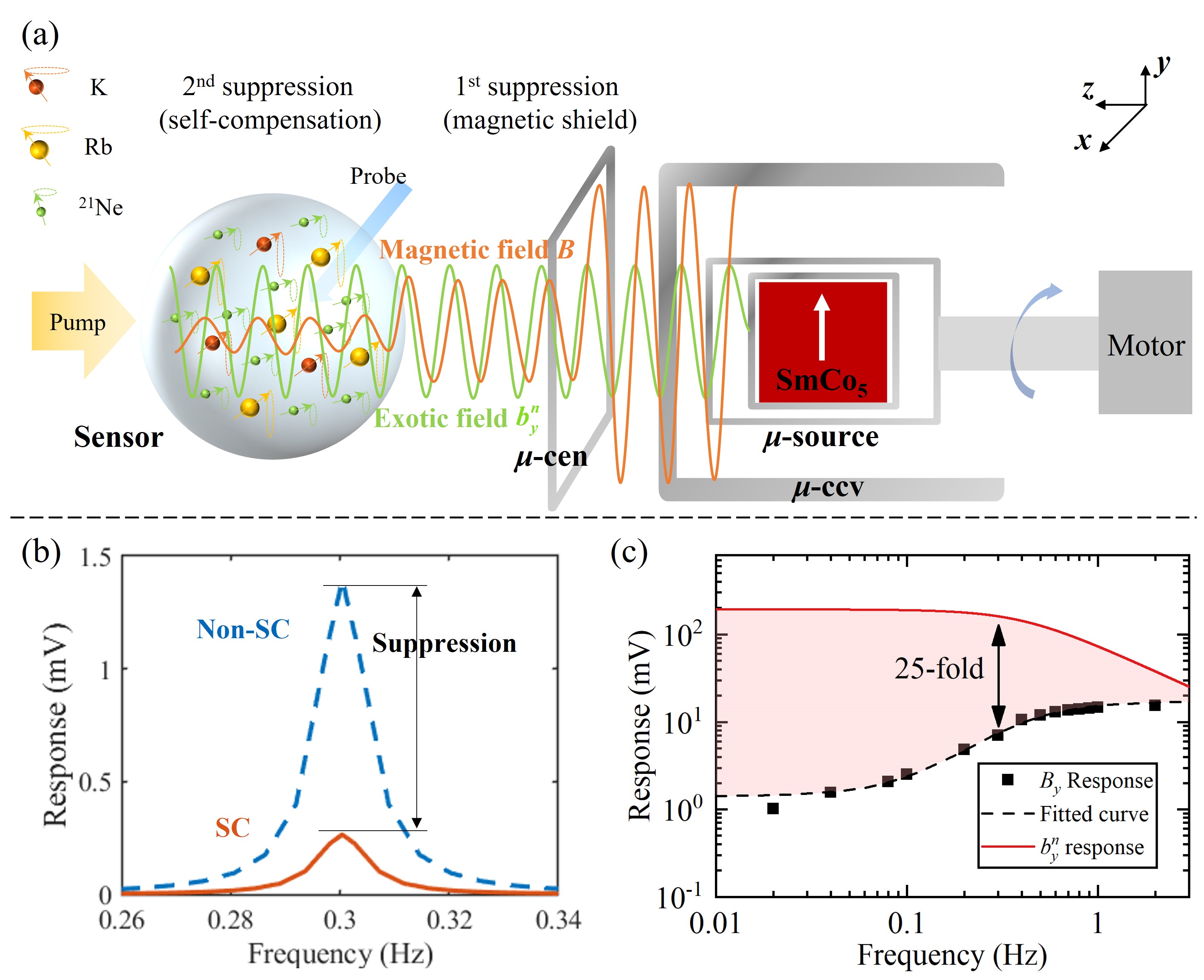}
\caption{ Principle of the experiment. (a) The atomic spins in the sensor cell are polarized and probed by pump and probe laser beams. The white arrow represents the magnetization direction of the spin source. A composite shield is used to shield the usual magnetic field from spin source, named as 1$^{\text{st}}$ suppression, which contains the $\mu$-source for enclosing the spin source, the $\mu$-ccv around the spin source and the $\mu$-cen for enclosing the sensor. The sensor works in the SC regime to provide 2$^{\text{nd}}$ suppression of magnetic field. Combining 1$^{\text{st}}$ and 2$^{\text{nd}}$ suppression, the magnetic field $B$ (orange line) is significantly suppressed while the sensitivity to exotic field $b_y^n$ (green line) is the highest. (b)  The measured response to magnetic field in SC regime is significantly suppressed against that out of SC. (c) In SC regime, the comagnetometer is insensitive to the low-frequency magnetic field while the sensitivity of exotic field is the highest. The response to usual magnetic field is suppressed by 25 times at 0.3\,Hz compared with the exotic field under the same condition. The measured frequency response is fitted with Bloch equations.}
\label{Fig.setup} 
\end{figure} 

The ISSC spin source consists of a cylindrical SmCo\(_5\) magnet with a diameter and height of 2.5\,mm with remanence of about 1\,T. The magnet is encased in a 2\,mm-thick layer of pure iron and a 2\,mm-thick layer of $\mu$-metal to minimize magnetic leakage. The core magnet and these shields together constitute what we define as the spin source. The primary advantage of the ISSC spin source is its combination of low-leakage magnetic field with a large net spin, due to the differential contributions of orbital and spin moments to the total magnetization \cite{ji2017searching}. After adding the electron spins from both the SmCo\(_5\) and the soft iron, the net number of spins is calculated to be $4.63\times 10^{20}$. The direction of the source spin is modulated by rotating it with a servo motor around the \textit{z}-axis, during which the spin remains perpendicular to the \textit{z}-axis. An exotic pseudomagnetic field shown in Eq.~\ref{eq.v3} that interacts with the atomic spins in the comagnetometer would generate a potentially detectable sinusoidal signal.
 
\begin{figure}[htb]
    \centering
\includegraphics[width= 0.95 \linewidth]{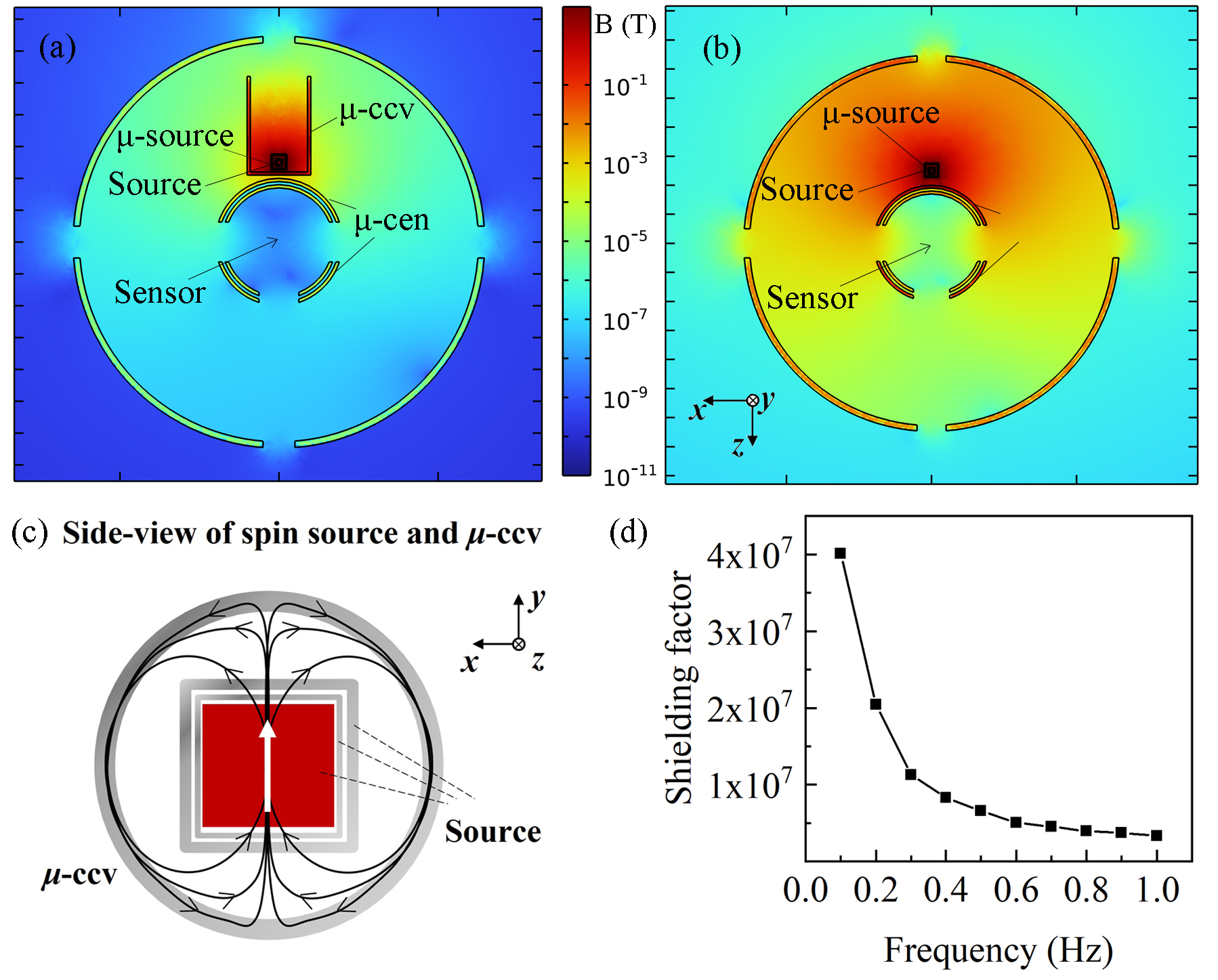}
\caption{Magnetic shield structure and performance. The magnetic shield structure (a) with the concave shield has much better performance than the structure (b) without the concave shield. \(\mu\)-source: shields of the spin source; \(\mu\)-ccv: concave shield; \(\mu\)-cen: central tape-wound shield.  (c) The $\mu$-ccv is specially designed to form a closed magnetic loop between $\mu$-ccv and the spin source, increasing the shielding factor by over two orders of magnitude to the unclosed condition. (d) The measured  shielding factor of the \(\mu\)-cen, \(\mu\)-ccv and the self-compensation property together is over \(1\times 10^7\) at the target frequency.}
\label{Fig.magshield} 
\end{figure}

These ISSC components are mounted on top of a long aluminum rod attached to a motor. The spin sensor is enclosed inside three layers of mu-metal magnetic shield ($\mu$-cen). To minimize the interaction distance and magnetic noise of shield,
these three layers of shield are made of a tape-wound magnetic shield, centered on the cell location, with a radius of 40 mm. To further enhance shielding performance against magnetic leakage from the source in a confined space, a special small-volume concave shield geometry ($\mu$-ccv) is applied. Fig.~\ref{Fig.setup} illustrates the schematic of the composite shielding, while the detailed geometry and shield performance are presented in Fig.~\ref{Fig.magshield}.

\textit{Magnetic field suppression --} As shown in Fig.~\ref{Fig.setup} (a), the K-Rb-\(^{21}\)Ne comagnetometer is operated in the self-compensation regime to suppress the influence of real magnetic fields. At the same time, the Fermi-contact interaction between $^{21}$Ne nuclei and alkali electrons increases the signal generated by the exotic field ~\cite{kornack2005nuclear,wei2023ultrasensitive,xu2022critical}. According to the Bloch equations~\cite{kornack2005nuclear, wei2022constraints} describing the coupled spin ensembles, the comagnetometer's response to quasi-static signal in the self-compensation regime is 

\begin{eqnarray}
P_{{\rm{x}}}^e = \frac{ P_z^e{\gamma ^e} R_2^e}{{{{R_2^e}^2} + {\gamma ^e}^2\delta {B_z}^2}} \left\{  {b_y^n - b_y^e}  + \frac{{\delta {B_z}}}{{{B_n}}} {B_y}\right\}\,, \label{eq:Bc signal}
\end{eqnarray}
where \(\gamma ^e\) 
is the gyromagnetic ratio of electron; \(P_z^e\) and \(R{{_2^e}}\) are the electron spin polarization and its transverse relaxation rate; 
\(\delta {B_z}=B_z-B_c\) is the difference between the bias field \(B_z\) and the self-compensation point \(B_c\). \(b_y^n\) and \(b_y^e\) are the exotic fields along \textit{y} axis that couple with nuclei and electrons. 
The spin polarization \(P_x^e\) is transformed to the output electric signal with a gain factor. 
When \(\delta {B_z}=0\), comagnetometer works in the SC regime where the response to usual magnetic fields $B_{y}$ is near zero. The mechanism behind SC regime is that the magnetization of noble-gas nuclear spins adiabatically follows the total magnetic field while its magnetization compensates $B_{y}$~\cite{PhysRevLett.89.253002,wei2023ultrasensitive}. We demonstrate the SC regime by comparing the measured responses to  $B_y$ in and out of SC regime by changing the bias field, as shown in Fig.~\ref{Fig.setup} (b). The response to $B_y$ is significantly suppressed in the SC regime. 

We use the same method as the Refs.~\cite{wei2023ultrasensitive,xu2022critical} to simulate the response to  the exotic field $b_y^n$ with the measured responses to the usual magnetic field $B_{y}$  based on the coupled Bloch equations. The responses to $b_y^n$ and $B_{y}$ are illustrated in Fig.~\ref{Fig.setup}(c). The SC regime works with frequency below 1\,Hz, where  the response to usual magnetic field is greatly suppressed compared with the response above 1\,Hz. Meanwhile, the response to the exotic field is the highest in SC regime, which is significantly higher than that of $B_{y}$. The suppression factor can be more than 10\(^4\) when the real magnetic field is DC.  To optimize the sensitivity to the spin-spin interaction, the spin source is rotated at the frequency where the response to the exotic field $b_y^n$ is high. Meanwhile,  to avoid the effect of the large residual usual magnetic field and the $1/f$ noise in low frequencies, we need to modulate the source at a  relatively higher frequency within the SC regime. Our compromise is to modulate the exotic field at  0.3\,Hz,
where the response to $B_{y}$ is suppressed by 25 times than the response of $b_y^n$.

The top view of the entire shielding structure is presented in Fig.~\ref{Fig.magshield}(a) and (b), showing simulations of the magnetic shielding effectiveness with and without the concave shield, $\mu$-ccv, respectively. The direction of magnetization of the spin source is perpendicular to the axial direction of the $\mu$-ccv. Thus a closed magnetic loop is formed by the side wall as shown in Fig.~\ref{Fig.magshield}(c). The inclusion of the concave shield enhances the shielding performance by more than 100 times within a limited space compared to the structure without the concave shape. 

The integrated magnetic suppression performance of the shield structure and the self-compensation ability is tested with an electromagnet with similar dimension as the source, and its amplitude and frequency is tuned by changing the current through it. The leakage of magnetic field can be estimated as $B^{l}_\text{cell} (f)= B^{l}_{ori}/\chi(f)$, where $B^{l}_{ori}$ is the leakage of the source which is measured to be approximately 1\,nT at a distance of 2\,cm, $\chi(f)$ is the decay factor of the magnetic leakage which includes the geometrical decay factor, the shielding factor of $\mu$-ccv and $\mu$-cen shielding,  and the suppression factor of the SC mode. Fig.~\ref{Fig.magshield} (d) shows the measured shielding factor of $\chi(f)$. At the working frequency of 0.3\,Hz, we calibrate the $\chi(0.3\,\rm{Hz})>1\times10^{7}$, and the magnetic leakage at the position of the sensor is estimated to be smaller than 0.1\,fT, which is confirmed by our experimental result as discussed below. It means that the leakage is about 16 orders of magnitude smaller than the magnet remanence of 1\,T. We can also consider the magnet do not have the composite shielding and estimate the magnetic leakage to be $10\mu$T at the position of the sensor, which means that our magnetic shielding together with the SC mode can effectively suppress the field by about 11 orders of magnitude.

\textit{Result --} The signal of comagnetometer and the pulse signal from  motor decoder, used for signal synchronization, are simultaneously acquired by a data acquisition device. The modulated exotic field applied on the sensor spins can be considered as an equivalent magnetic field by integrating the overall contributions from the spins in the source with $b^{n,p}=\eta^{n,p}/\mu_N\int V_{pp} dV$, where $\eta^n=0.58$ and $\eta^p=0.04$ are the fraction factors for neutron and proton spin polarization, while $\mu_N$ the magnetic moment of the $^{21}$Ne nucleus. The detailed data processing procedure and the evaluation of the phase uncertainty, as summarized in Table\,\ref{tab:errors} are provided in the Appendix.

\begin{figure*}[htb]
    \centering
\includegraphics[width= 0.99 \linewidth]{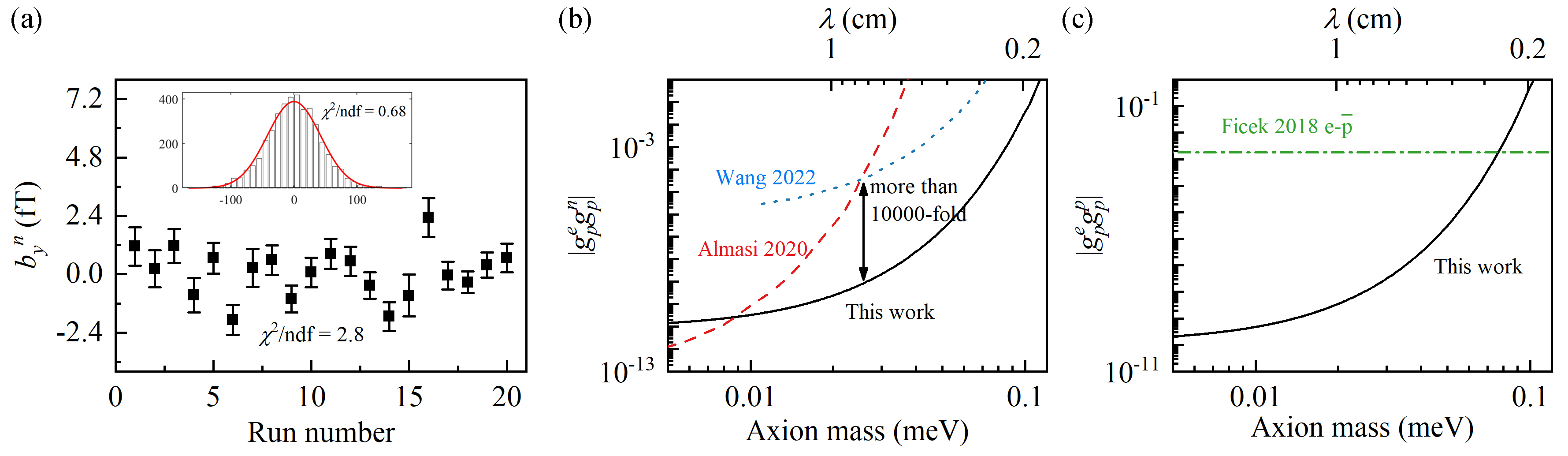}
\caption{Experimental results. (a) Each point represents an average of a 4\,h measurement. The error bars represent the statistical and systematic error of the comagnetometer. The exotic field \(b_y^n\) is measured to be \((0.08 \pm 0.22_{\text{stat}})\)\,fT. Inset: histogram of a 4\,h dataset with a Gaussian fit in red line. (b) The experimental limits on \(|g_p^e g_p^n|\). The blue dotted line, ``Wang 2022", is from Ref.\,\cite{wang2022limits}, the red dashed line, “Almasi 2020", is from Ref.\,\cite{almasi2020new}. Almasi 2020 is more sensitive at low axion masses due to a larger spin source. The main gain over Wang 2022 is the use of a higher-density (solid) spin source and a higher-sensitivity sensor in this work. (c) The experimental limits on \(|g_p^e g_p^p|\). The green dashed-dotted line, 'Ficek 2018 e-\(\bar{\text p} \)', is from Ref.~\cite{PhysRevLett.120.183002}.
}
\label{Fig.result} 
\end{figure*}

Using all the 80-hour data, the exotic field \(b_y^n\) is measured to be \((0.08 \pm 0.22_{\text{stat}})\)\,fT as shown in  Fig.~\ref{Fig.result} (a). At a force range of 7\,mm (equivalent to axion mass of about $0.03\,\rm{meV}$), since no significant evidence for the axion-mediated force is found, taking into account various systematic uncertainties  shown in Table\,\ref{tab:errors} as well as the statistical uncertainty, new limit on the dimensionless coupling factor of the new force are obtained.
The coupling constant is determined to be \(g_p^eg_p^n= (0.38\pm 1.08_\text{stat}\pm0.05_\text{sys} )\times 10^{-9}\), and we set a limit \(|g_p^eg_p^n| \leqslant 2.5 \times 10^{-9}\) with 95\% confidence level, improving on the previous limit~\cite{wang2022limits} by over four orders of magnitude. The limit for other force range are determined with the same method, as shown in Fig.~\ref{Fig.result} (b).
Furthermore, taking the contribution of the proton spin in $^{21}$Ne nucleus into account, we establish the strongest constraints on the coupling between electrons and protons, as shown in Fig.~\ref{Fig.result} (c). There is no reported electron proton coupling in this range, and we include the electron antiproton coupling from antiprotonic helium result as a comparison 
\cite{PhysRevLett.120.183002}.  
Constraints between electron and proton interactions also exist at shorter force range \cite{fadeev2022pseudovector,cong2024improved}. A NMR resonance enhancement experiment is also proposed to detect axions within this force range \cite{arvanitaki2014resonantly}.

\begin{table}[htb]
\centering
\caption{The experimental parameters and their error budget for the coupling coefficients.}
\begin{tabular}{ccc}
    \hline
    \hline
    Parameter & Value  &  \(\Delta g_p^eg_p^n (\times 10^{-10})\)\\ 
    \hline
    {\multirow{2}{*}{Net spin (\(\times 10^{20}\))}} & {\multirow{2}{*}{\(4.63(15)\)}} & -0.12\\
    {} & {} & +0.13\\
    \hline
    \multirow{2}{*}{\(K_{b_y^n}^\text{AC}\) ($\mu$V/fT)} & \multirow{2}{*}{\(1.09(10)\)} & -0.32\\ 
    {}&{} &+0.38\\
    \hline
    Mounting error (deg) &\(<2.00\) &\(<0.18\)\\ 
    \hline
    \(r\) (mm) & \(50.00(5)\)  &\(\pm 0.04\)\\
    \hline
    Phase uncertainty (deg) &\(\pm 1.57\) &\(<0.15\)\\
    \hline
        \multicolumn{2}{c}{Total}  & \(<0.52\)\\
    \hline
    \hline
\end{tabular}
\label{tab:errors}
    \end{table}

The main systematic uncertainties arise from experimental setup and physics simulations, including position uncertainties, the phase uncertainty of the comagnetometer, and net spin simulation errors and the mounting error (misalignment) of the spin source. The long-term stability of the comagnetometer was monitored over a period of more than 24 hours, revealing that the fluctuation of the key parameter, the self-compensation point, was less than 0.3\% of its mean value. However, drift in the self-compensation point introduces errors in the calibration factor $K_{b_y^n}^\text{AC}$. These uncertainties are summarized in Table~\ref{tab:errors}.

Future improvements will focus on detecting exotic forces at shorter ranges with higher sensitivity. The primary challenge remains the trade-off between reducing magnetic leakage and increasing the density of the spin source. In this work, we addressed this issue by designing a specially structured shield to create a magnetic loop, enhancing shielding performance by 100-fold within a compact volume. Further advancements require shields made of materials with higher permeability and lower noise to achieve better suppression of magnetic leakage and further improve sensitivity.

Furthermore, our approach offers broader possibilities for applications requiring both high magnetic field regions and ultra-low magnetic field regions. As a case study, we focus on the zero- to ultralow-field nuclear magnetic resonance (ZULF-NMR) technique in biological and chemical research~\cite{tayler2017invited,jiang2021zero}. ZULF-NMR necessitates spatially distinct areas: a tesla-level magnetic field region for prepolarization and an ultralow magnetic field region (at the femtotesla level) for NMR signal detection. Conventional setups employ cylindrical magnetic shielding and maintain a significant gap (approximately 40 cm) between the high- and low-field regions, introducing notable signal attenuation due to relaxation effects~\cite{ledbetter2009optical,tayler2017invited,jiang2018experimental,jiang2019magnetic}. However, the small concave shielding structure presented here reduces the sample transportation time by up to an order of magnitude and reduces the setup volume by over 15-fold. For ZULF-NMR experiments, the shielding factor of the $\mu$-source and $\mu$-ccv is improved to $10^8$, even with a 6-mm diameter axial hole for sample shuttling. The reduction in distance and time enabled by our design provides significant advantages in preserving signal integrity and simplifying the experimental setup.

\textit{Conclusion --} In conclusion, we utilized a comagnetometer and a specially designed spin source to search for exotic dipole-dipole interaction. The usual magnetic field from the source is suppressed by over 11 orders of magnitude, due to source and sensor shielding and the use of the self-compensation regime, while maintaining the ultrahigh sensitivity to the exotic field. With the high spin density of the  source, we have improved the limit on spin-dependent interaction between electrons and neutrons mediated by axion by over 10000 times. We also set  the strongest constraints on the coupling between electrons and protons, which previously have only been reported through analysis of spectroscopy data from antiprotonic helium  
\cite{PhysRevLett.120.183002}. In future work, the experimental methods developed here will be applied to a broader search for exotic spin-dependent interactions, search for dark matter directly coupled to atomic spins, and study of spin-gravity interaction. New limits can be set on coupling parameters for spin-1 bosons \cite{cong2024spin}, such as the \( Z' \) boson, and also for study of violation of parity symmetry, as will be discussed in Ref.\,\cite{xu2024preparation}.

\section*{Acknowledgement}

K. W. was funded by the National Science Foundation of China (NSFC) under Grants No. 62203030 and 61925301 for Distinguished Young Scholars, by the Innovation Program for Quantum Science and Technology under Grant 2021ZD0300401, by the Fundamental Research Founds for the Central Universities. W. J. and D. B. was funded by the DFG Project ID 390831469: EXC 2118 (PRISMA+ Cluster of Excellence), by the COST Action within the project COSMIC WISPers (Grant No. CA21106), and by the QuantERA project
LEMAQUME (DFG Project No. 500314265). 

\endgroup

\end{thebibliography}

\clearpage 
\section*{Appendix}
\label{sec:endmatter}

\subsection*{Data Processing}
\begin{figure}[htb]
    \centering
\includegraphics[width= 0.99 \linewidth]{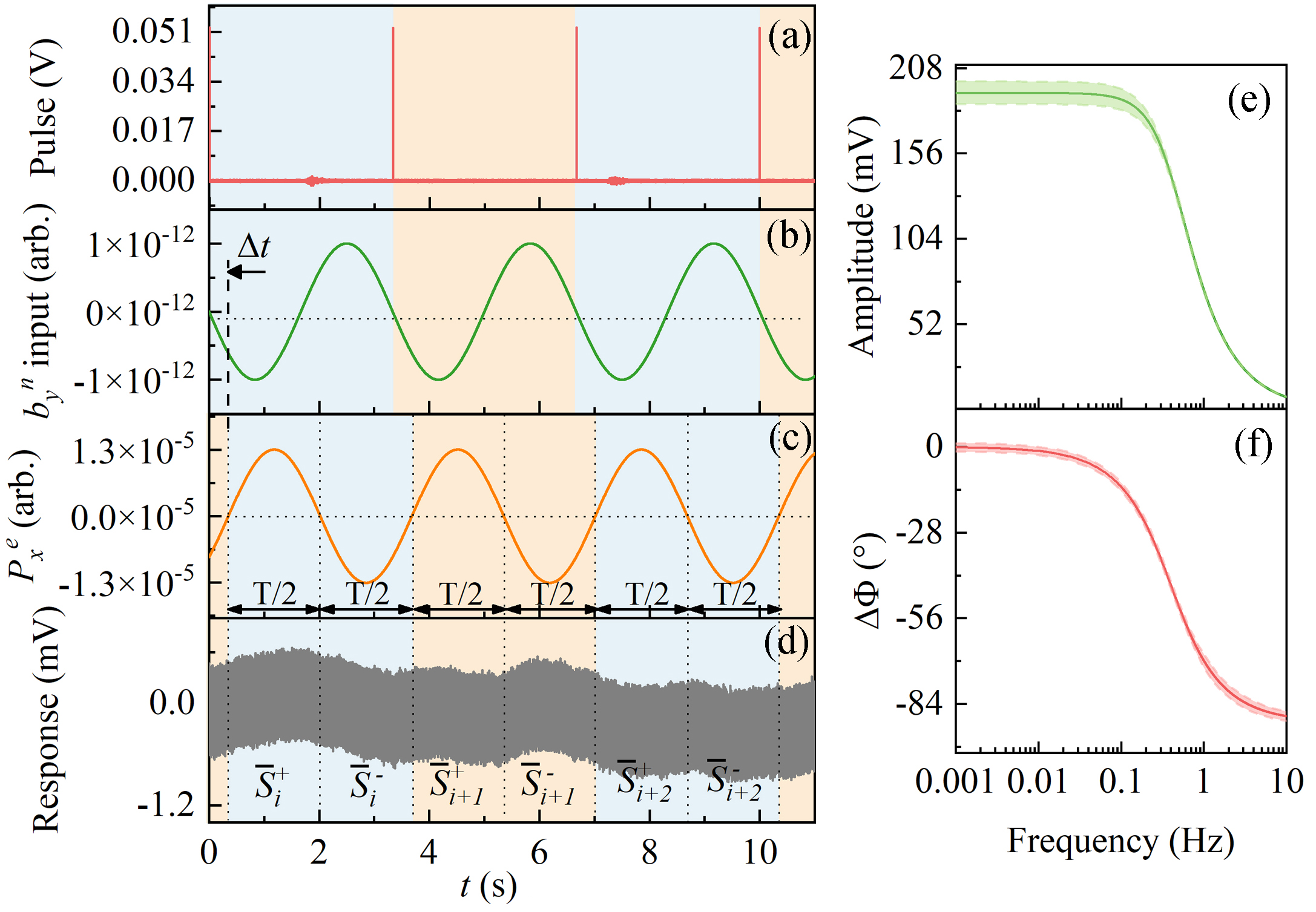}
\caption{Data acquisition process. (a) The motor decoder generates one pulse per revolution. (b) As the motor rotates, the projection of the exotic field \(b_y^n\) on the \textit{y}-axis produces an effective sinusoidal magnetic field. (c) The simulated response of the comagnetometer shifted is phase relative to the input. (d) The measured response of the comagnetometer. The simulated (e) amplitude and (f) phase response of the comagnetometer to \(b_y^n\) in the self-compensation regime. 
The uncertainty bands are calculated based on the uncertainties of the measured parameters in the text. }
\label{Fig.experiment} 
\end{figure}

 The angular position of the rotating spin source is indicated by the pulse in Fig.~\ref{Fig.experiment} (a). The projection of exotic field along the sensitive axis  $b_y^n$ is approximately a sinusoidal field, see Fig.~\ref{Fig.experiment} (b). The response of the comagnetometer to $b_y^n$ is simulated based on measured parameters, and is shown in Fig.~\ref{Fig.experiment} (c). The corresponding experimental data from the comagnetometer is shown in Fig.~\ref{Fig.experiment} (d).  The responses to the exotic field can be divided into multiple periods  based on the pulse marker and phase shift. A slight drift is observed in the experimental data, attributed to air convection disturbances and temperature fluctuations.

As shown in Fig.~\ref{Fig.experiment} (d), the data are divided into segments, with each segment corresponding to one modulation period. Using linear interpolation between consecutive pulses, each segment is further divided into two parts, corresponding to two half-periods. Then the mean value of each segment is derived and denoted as $\overline{S}_i^+$ and $\overline{S}_i^-$ for the positive and negative parts, respectively. Assuming that the background signal is time dependent, we can use the following procedure to remove the DC bias field and low-frequency drifts and get the effective amplitude of the signal for the i-th period of the data~\cite{wei2022constraints}:

\begin{equation}
    \overline{S}_i=\frac{1}{8} [\overline{S}_{i}^+ -3 \overline{S}_{i}^- + 3 \overline{S}_{i+1}^+ -\overline{S}_{i+1}^-]\,.
\end{equation}

We further consider the phase shift between the sensor response $S$ and the exotic field $b^n_y$. In Fig.~\ref{Fig.experiment} (f), the phase shift is calculated based on the measured parameters \(R_{2n}\)=0.030(2) rad/s, \(R_{2e}\)=5430(200) rad/s, \(P_z^e\) =0.75(11), \(B_c\)=506(1) nT, \(B_z^n\)=407(7) nT and \(B_z^e\)=100(7) nT \cite{wei2022constraints}. The shaded area of curve is the corresponding uncertainty. The phase shift  is \(\Delta \Phi = -36.3(16)^\circ\) at modulation frequency 0.3\,Hz  (corresponding to a time delay of \(\Delta t = 0.33(2)\)\,s). The signal segments shown in Fig.~\ref{Fig.experiment} (c) and (d) is shifted correspondingly by $\Delta t$. The  exotic field $b_y^n$ is derived from sensor signal $S$ by $S= K_{b_y^n} b_y^n$. The calibration factor $K_{b_y^n}$ of $b_y^n$ is determined by the measured response to usual magnetic field ~\cite{kornack2005nuclear,wei2022constraints,PhysRevResearch.6.013339}, as shown in Fig.~\ref{Fig.experiment} (e). Moreover, we consider the difference between the DC and AC response factor with a ratio $K_{b_y^n}^\text{AC}/ K_{b_y^n}^\text{DC}$, for 0.3\,Hz AC signal, which is determined to be 0.84(6).
\end{document}